\documentclass[a4paper,twosides]{article}
\usepackage{CJK,multicol,multirow,graphics,fancyhdr,epstopdf}
\usepackage{amsmath,amsfonts,amssymb,bm,upgreek,mathrsfs,mathcomp}
\usepackage[pagewise,switch,columnwise]{lineno}
\usepackage[compress,nospace]{cite}
\usepackage[dvipsnames]{xcolor}
\usepackage{wasysym}

\usepackage[colorlinks,linkcolor=blue,urlcolor=blue,citecolor=blue,anchorcolor=blue,pdfborder=000,
 pdftitle={Changepdftitle}, pdfauthor={Changepdfauthor},
 pdfcreator=CHIN. PHYS. LETT., pdfproducer=CPL]{hyperref}

\newcommand{\ucite}[1]{$\!^\text{\cite{#1}}$} 
\definecolor{blue}{RGB}{46,48,146}

\def\headrule{\kern 1mm \hrule width 17cm \kern -1mm}%
\def\footnoterule{\kern 1mm \hrule width 7cm \kern 2.2mm}%
\newcommand{\figcaption}[3]{\centerline{\footnotesize \begin{tabular}{p{#1cm}}{\bf Fig.~{#2}. }#3\end{tabular}}} 

\newcommand{\fref}[2]{\hyperlink{f#1}{#2}} 
\newcommand{\fl}[1]{\hypertarget{f#1}{}} 
\newcommand{\el}[1]{\label{e#1}}   

\newcommand{\cplyear}{2024} \newcommand{\cplvol}{41}
\newcommand{\cplno}{10} \newcommand{\cplpagenumber}{107503}
 \newcommand{\cplpage}{\cplpagenumber-\thepage}

\begin{document}

\begin{CJK}{GBK}{song}\vspace* {-4mm} \begin{center}
\large\bf{\boldmath{Magnetism Measurements of Two-Dimensional van der Waals Antiferromagnet CrPS$_{4}$ Using Dynamic Cantilever Magnetometry}}
\footnotetext{\hspace*{-5.4mm}\noindent$^{*}$Corresponding authors. Email: kwang@hmfl.ac.cn; lkzou2023@sinano.ac.cn; zhangjinglei@hmfl.ac.cn


\noindent\copyright\,{\cplyear}
\href{http://www.cps-net.org.cn}{Chinese Physical Society} and
\href{http://www.iop.org}{IOP Publishing Ltd}}
\\[6mm]
\normalsize \rm{}Qi Li$^{1,2}$, Weili Zhen$^{1}$, Ning Wang$^{1}$, Meng Shi$^{1,2}$,\\ Yang Yu$^{3}$, Senyang Pan$^{1,2}$, Lin Deng$^{1,2}$, Jiaqiang Cai$^{1,2}$,\\ Kang Wang$^{1*}$, Lvkuan Zou$^{4*}$, Zhongming Zeng$^{4}$,\\ Zhaosheng Wang$^{1}$, and Jinglei Zhang$^{1*}$
\\[3mm]\small\sl $^{1}$Anhui Province Key Laboratory of Condensed Matter Physics at Extreme Conditions, High Magnetic Field Laboratory, HFIPS, Chinese Academy of Sciences, Hefei 230031, China

$^{2}$Science Island Branch of Graduate School, University of Science and Technology of China, Hefei 230026, China

$^{3}$School of Advanced Manufacturing Engineering, Hefei University, Hefei 230601, China

$^{4}$Suzhou Institute of Nano-tech and Nano-bionics, Chinese Academy of Sciences, Suzhou 215123, China
\\[4mm]\normalsize\rm{}(Received 2 September 2024; accepted manuscript online 23 September 2024)
\end{center}
\end{CJK}
\vskip 1.5mm

\small{\narrower Recent experimental and theoretical work has focused on two-dimensional van der Waals (2D vdW) magnets due to their potential applications in sensing and spintronics devises. In measurements of these emerging materials, conventional magnetometry often encounters challenges in characterizing the magnetic properties of small-sized vdW materials, especially for antiferromagnets with nearly compensated magnetic moments. Here, we investigate the magnetism of 2D antiferromagnet CrPS$_{4}$ with a thickness of 8\,nm by using dynamic cantilever magnetometry (DCM). Through a combination of DCM experiment and the calculation based on a Stoner--Wohlfarth-type model, we unravel the magnetization states in 2D CrPS$_{4}$ antiferromagnet. In the case of $H\parallel c$, a two-stage phase transition is observed. For $H\perp c$, a hump in the effective magnetic restoring force is noted, which implies the presence of spin reorientation as temperature increases. These results demonstrate the benefits of DCM for studying magnetism of 2D magnets.

\par}\vskip 3mm
\noindent{\narrower{DOI: \href{http://dx.doi.org/10.1088/0256-307X/\cplvol/\cplno/\cplpagenumber}{10.1088/0256-307X/\cplvol/\cplno/\cplpagenumber}}

\par}\vskip 5mm
\begin{multicols}{2}

Recent years, two-dimensional (2D) magnetism in layered van der Waals (vdW) materials has attracted considerable attention since potential applications in 2D spintronics.\ucite{1,2,3,4,5,6,7} Directly measuring magnetization is crucial to revealing the magnetic structures underlying nontrivial behaviors,\ucite{8,9,10} such as abnormal magnetizing,\ucite{11} magnetic phase transitions,\ucite{12,13,14} and hidden orders.\ucite{15,16} However, conventional magnetic measurement techniques encounter difficulties when probing the magnetic properties of nano-scale vdW magnets. For example, vibrating sample magnetometry has a sensitivity of $10^{-9}$\,A$\cdot$m$^2$, which is insufficient for such studies.
	
Dynamic cantilever magnetometry (DCM) has emerged as an ultra-sensitive technique,\ucite{17,18,19} the method enables the detection of magnetic moment signals at the magnitude of $10^{-17}$\,A$\cdot$m$^2$.\ucite{20} When a nanometer-sized sample is attached onto the free end of a cantilever, the effective spring constant of cantilever alters with external magnetic field due to magnetic restoring force exerting on the sample. This additional magnetic spring constant $\Delta k$ is associated with the magnetization of the sample.\ucite{21} Based on this relationship, the magnetic properties of sample can be derived from eigen frequency shifts $\Delta f$ of the cantilever sensor. So far, DCM has been employed in measurements of individual nanometer-scaled particles such as magnetic nanotube and magnetic mesocrystals,\ucite{18,22,23,24,25,26} which can be mounted onto cantilever by using micro-manipulator. However, due to the lack of appropriate transfer technique, the thickness of vdW materials studied by DCM has been limited to micrometer scale.\ucite{27}
	
In this work, we develop a two-step fabricating method for transferring exfoliated atomically thin flakes onto cantilever sensor, enabling us to extend DCM measurements to 2D magnets. Employing this technique, we investigate the magnetic properties in thin CrPS$_{4}$ flake with a thickness down to 8\,nm by measuring the $\Delta k$. In the case of $H\parallel c$, the spin flop transition leads to a dip in $\Delta k$, followed by a gradual increase and then reaches next region with a change in slope. For the case of $H\perp c$, the low-temperature results indicate a canted antiferromagnetic state. As $T$ increases, a bump followed by a hump in $\Delta k$ is observed until $\Delta k$ is completely negative above 30\,K, which can be attributed to the presence of spin reorientation. The $H$--$T$ phase diagrams of 2D CrPS$_{4}$ revealed by DCM experiment are consistent with the previous studies on bulk samples,\ucite{28} which demonstrates the reliability of DCM in investigating magnetic properties of 2D magnets.

CrPS$_{4}$, a ternary transition metal chalcogenide (TTMC), has monoclinic symmetry with space group $C2/m$.\ucite{29,30} Its closely packed surface is constituted by S atoms in monolayers as presented in Fig.\,\fref{1}{1(a)}. Magnetically, CrPS$_{4}$ exhibits intralayer ferromagnetism and interlayer antiferro-coupling,\ucite{31,32} which contribute to a complex metamagnetic phase transition. The weak spin interactions in CrPS$_{4}$ leads to little dependence of transition temperature on thickness.\ucite{32} Thus, CrPS$_{4}$ is an appropriate candidate for demonstrating the capability of DCM in examining the magnetism in 2D magnets. The zero-field-cooled (ZFC) magnetization characterizations of bulk CrPS$_{4}$ were measured with external field $\mu _0 H =0.1$\,T applied both parallel to the $c$ direction [Fig.\,\fref{1}{1(b)}] and in the $ab$ plane [Fig.\,\fref{1}{1(c)}]. The N\'eel temperature of our sample is around 38\,K, which is consistent with previous studies.\ucite{29}

\end{multicols}

\fl{1}\centerline{\includegraphics{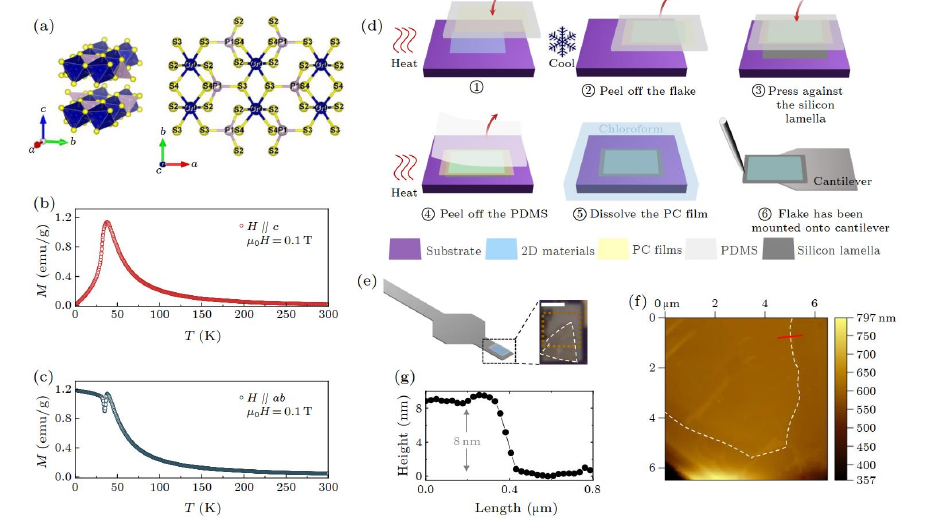}}

\vskip 4mm

\figcaption{15}{1}{Characterization of CrPS$_{4}$ and fabrication of the DCM device. (a) The crystal structure of CrPS$_{4}$. [(b), (c)] The $c$-direction and $ab$-plane magnetic susceptibilities, respectively, measured under a magnetic field of ${\mu _0} H=0.1$\,T. (d) Schematic representation of the transferring method of the 2D CrPS$_{4}$ flakes. Gray needles represent the micro manipulators for transfer. The main steps have been arranged in sequential order using numbered circles. (1) The polycarbonate (PC) film was affixed onto the polydimethylsiloxane (PDMS) and pressed against the flake, followed by heat melting to enhance adhesion. (2) Cool the substrate to solidify the PC film and peel off the flake. (3) The flake was aligned on top of the target silicon lamella and pressed against the lamella. (4) Heat the substrate and peel off the PDMS. (5) Soak the substrate in chloroform and dissolve the PC film. (6) Dry off the substrate and transfer lamella to the cantilever sensor. (e) Schematic drawing of the DCM devices S1. The inset shows the photograph of the silicon lamella with sample attached on (scale bar: 5\,$\upmu$m). (f) Atomic force microscopy image of the flake on S1 with a thickness of 8\,nm. The thickness of the flake is shown in (g).}

\begin{multicols}{2}
	
We designed and fabricated sensitive cantilever beam for DCM measurements. Firstly, a thin film of SiO$_{2}$ with thickness of 1\,$\upmu$m was grown on top Si layer of a silicon-on-insulator (SOI) substrate by plasma-enhanced chemical vapor deposition (PECVD) to balance stress beneath. Patterns of cantilevers were micro-fabricated by ultra-violet lithography on top surface of SOI. An inductively coupled plasma (ICP) deep reactive ion etch (DRIE) was adopted to etch SOI to buried oxide (BOX) layer. On bottom surface of SOI substrate, the patterned 500\,$\upmu$m silicon was removed by DRIE (Bosch Process) on a deep silicon etcher system. Finally, sacrificial SiO$_{2}$ layer was released by dry release etch using vapor HF (hydrogen fluoride) chemistries (sacrificial vapor release etching).

Conventionally, a focused ion beam (FIB) is used to structure micro/nano-sized samples and transfer them onto the cantilever,\ucite{18} especially for thin films.\ucite{19} For 2D flakes, FIB fabricating might induce damage in samples. In our study, a two-step transfer method is schematically illustrated in Fig.\,\fref{1}{1(d)}, whose process comprises two essential steps. Initially, the exfoliated 2D flake was mechanically peeled off and transferred onto a silicon lamella. Subsequently, by using the micro-manipulators, the 2D flake together with the lamella were mounted onto the free end of a cantilever. This technique provides two advantages in DCM experiment: (1) a major reduction in thickness, (2) a major boost in surface area for thin film to enhance the magnetic signal. The lamella attached to the extremity of cantilever can play a role as a mass-loaded end to enhance the sensitivity. Moreover, the use of lamella facilitates the mounting process. By utilizing this technique, we have successfully fabricated CrPS$_{4}$ sample S1 with a thickness down to 8\,nm and an area of 30\,$\upmu$m$^2$. The silicon lamella was cut to a size of $12\,\upmu$m\,$\times 8\,\upmu$m $\times 1\,\upmu$m by using FIB. Its magnetic signal can be negligible in DCM experiments. The single-crystal Si cantilever used in our study is 200\,$\upmu$m long, 10\,$\upmu$m wide and 200\,nm thick. The morphological characterization depicted in Fig.\,\fref{1}{1(e)} demonstrates that our method is a clean transfer technique.

\vskip 4mm

\fl{2}\centerline{\includegraphics{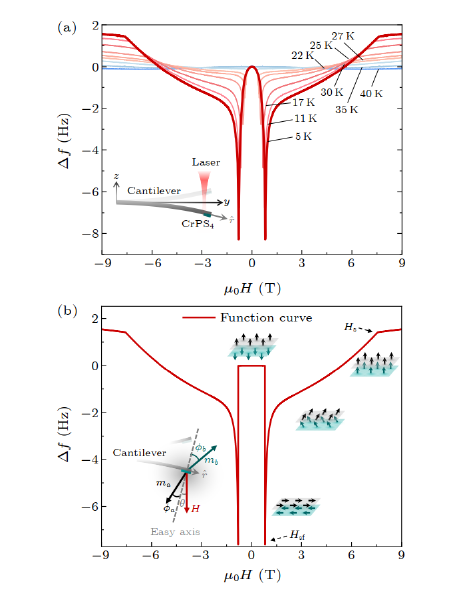}}

\vskip 2mm

\figcaption{7.5}{2}{The experimental and calculated results for the magnetic field dependence of $\Delta f$ in the case of $H\parallel c$. (a) The field dependence of $\Delta f$ in S1 with $H\parallel c$ from 5\,K to 40\,K. Inset schematically illustrates the experimental setup. The cantilever is securely fixed along the $y$-axis and vibrates within the $y$--$z$ plane. The laser is directed along the $z$-axis. (b) The calculated $\Delta f$ for an antiferromagnetic thin film when $H$ is applied along the easy axis. The red line illustrates the numerical results. The left inset exhibits the schematic view of the relative orientations of the cantilever axis, applied magnetic field $H$, and the magnetic moments in sample. Other insets depict the deduced spin alignments of CrPS$_{4}$ as black and green arrows.}

\medskip

The DCM insert was sealed in a vacuum sleeve, maintaining a pressure below $10^{-6}$\,mbar. By using a rotatable sample stage, a magnetic field up to 9\,T can be applied parallel or perpendicular to the cantilever. The cantilever used in our DCM has a low spring constant of 0.14\,mN/m. The eigen frequency $f_0$ of 2.7\,kHz can be measured by using a home-made laser interferometer. The frequency responses were precisely monitored by employing the ring-down method. When the signal is strong, due to the mechanical properties of the cantilever beam, the antisymmetric effect in magnetic fields is obvious. In order to exclude the antisymmetric effect in magnetic fields, the $\Delta f$ has been symmetrized in positive and negative magnetic fields.
	
Figure~\fref{2}{2} displays the DCM experiment results in S1 for $H\parallel c$. The measured $\Delta f$ as a function of $H$ at low temperatures could be classified into three regimes. In the low-field range below a critical field of around 0.8\,T, $\Delta f$ rapidly decreases to a minimum. In the intermediate regime, $\Delta f$ gradually increases and then enters next region at around 7.6\,T. The similar behavior has typically been observed in magnetic torque signals of bulk CrPS$_{4}$,\ucite{29} however, its quantitative interpretation of magnetism is still missing.
	
In order to analyze the magnetic phase transitions at low temperatures, we treated the sample as a uniaxial particle by following the Stoner--Wohlfarth-type model presented by Weber {\it et al.}\ucite{18} For CrPS$_{4}$ samples, as reported previously, there are two inequivalent magnetic structures that induce complicated anisotropies.\ucite{30} Nevertheless, at low temperature, the out-of-plane anisotropy dominates. Thereby we considered a projected anisotropy constant $K$ ($K>0$) to simplify the analysis. In this model, two sets of macrospins ($m_a$ and $m_b$) with interactions have been presented. The spin alignments have been depicted as green and black arrows in Fig.\,\fref{2}{2(b)}. The $\phi_a$ and $\phi_b$ denote the deviations of $m_a$ and $m_b$ from the easy axis. The two sets of macrospins remain in equilibrium between the inherent anisotropy, magnetic torque induced by $H$ and the antiferromagnetic interactions. An antiferromagnetic interaction constant $A$ has been introduced. $Am_a$ can be regarded as an effective magnetic field applied to $m_b$. Thus, the free energy of the system can be written as the sum of the cantilever energy, the Zeeman energy, an effective anisotropy energy and an antiferromagnetic interaction energy
\begin{align}\el{1}
E={}&\frac{1}{2} k_0 l_{\rm e}^2 \theta^2-(M_a V H \cos (\phi_a-\theta)\notag\\
&+M_b V H \cos (\phi_b-\theta))\notag\\
&+\frac{1}{2} K V (\sin ^2 \phi_a+\sin ^2 \phi_b)\notag\\
& +A M_a M_b V^2 \cos (\phi_a-\phi_b),
\end{align}
where $k_0$ and $l_{\rm e}$ are the inherent spring constant and effective length of the cantilever, respectively. $V$ is the volume of the CrPS$_{4}$ sample. The magnetizations of two macrospins are $M_a$ and $M_b$ with $M_a=M_b=M_0$. As shown in inset of Fig.\,\fref{2}{2(b)}, $\theta$ depicts the vibration deviation. Considering the system in an equilibrium state, the energy of the system must satisfy $\frac{\partial E}{\partial \phi_a}=\frac{\partial E}{\partial \phi_b}=0$, giving
\begin{align}\el{2}
&\frac{\partial E}{\partial \phi_a}=M_a V H \sin (\phi_a-\theta)+\frac{1}{2} K V \sin 2 \phi_a\notag\\
&-A M_a M_b V^2 \sin (\phi_a-\phi_b)=0,\notag \\
&\frac{\partial E}{\partial \phi_b}=M_b V H \sin (\phi_b-\theta)+\frac{1}{2} K V \sin 2 \phi_b\notag\\
&+A M_a M_b V^2 \sin (\phi_a-\phi_b)=0.
\end{align}
	
The solutions of the equation reveal the possible spin alignments in samples. One solution for $\theta=0$ is $\phi_{a_0}=-\phi_{b_0}=0$, which indicates an A-type antiferromagnetic (A-AFM) spin arrangement. When $H$ reaches a critical value $H_{\rm sf}$, which satisfies $M_0 H_{\rm sf}+(K-A V M_0^2)(\frac{2 M_0 H_{\rm sf}}{-K+2 AV M_0^2}- \frac{-K+2 AV M_0^2}{M_0 H_{\rm sf}^2})=0$, the macrospins prefer to stay in plane with lower total energy and the $H_{\rm sf}$ was denoted as spin-flop field. Correspondingly, the macrospins align as $\phi_{a_0}=-\phi_{b_0}= \arccos (\frac{M_0H}{-K+2AM_0^2V})$, revealing the canted spin alignments. With increasing $H$, the macrospins gradually rotate to the external field and orient towards the same direction at $H_{\rm s}=\frac{-K+2 A M_0{}^2 V}{M_0}$. The connection between the spin alignment and the mechanical response of the cantilever sensor was established by the formula $\Delta k=\frac{1}{l_{\rm e}^2}\frac{\partial^2 E}{\partial \theta^2}-k_0$ (see chapter S4 in the Supporting Information for more details). According to $\Delta f=f_0 \frac{\Delta k}{2k_0}$, we can obtain the calculated $\Delta f$ as a function of $H$,
\begin{align}\el{3}
\Delta f={}&\frac{f_0 V}{2 k_0 l_{\rm e}^2}\Big(\frac{H K M_0}{K+H M_0+A M_0^2 V}\nonumber\\
&-\frac{H K M_0}{K-H M_0+A M_0^2 V}\Big), ~{\rm for}~|H|<H_{\rm sf}, \nonumber\\
\Delta f={}&\frac{f_0 V}{2 k_0 l_{\rm e}^2} \cdot 2 M_0 H K\Big(\frac{2 H^2 M_0^2}{(-K+2 A V M_0{}^2)^2}-1\Big)\nonumber\\
&\Big[M_0 H+(K-A V M_0^2)\Big(\frac{2 M_0 H}{-K+2 AV M_0^2}\nonumber\\
&-\frac{-K+2 AVM_0^2}{M_0 H^2}\Big)\Big]^{-1},~{\rm for}~H_{\rm sf}<|H|<H_{\rm s},\nonumber\\
\Delta f={}&\frac{f_0 V}{2 k_0 l_{\rm e}^2}\frac{2 M_0 |H| K}{M_0 |H|+K-A V M_0^2},~{\rm for}~|H|>H_{\rm s}.
\end{align}

\vskip 4mm

\fl{3}\centerline{\includegraphics{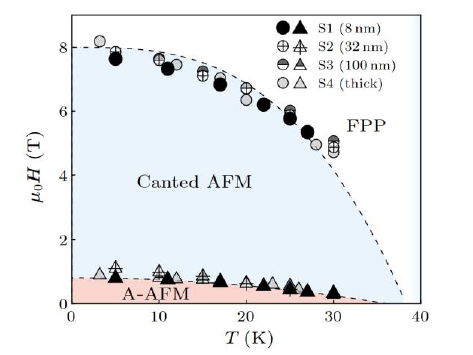}}

\vskip 2mm

\figcaption{7.5}{3}{Magnetic phase diagram of CrPS$_{4}$ determined from the DCM measurements with $H$ applied along $c$ axis. The spin flop boundaries are represented as triangles (circles for the saturation states) and four patterns denote four different samples.}

\medskip

The numerical results have been shown in Fig.\,\fref{2}{2(b)}, which are consistent with the experiment data. The singularities at $H= \pm H_{\rm sf}$ reflects breakdown in $\Delta f$--$H$ curve. The corresponding $\phi_{a_0}$ and $\phi_{b_0}$ indicate that the macrospins, initially aligning as an A-AFM spin alignment, flop down into the $ab$ plane. By combining the calculated $\Delta f$ and the experiment results, the dips in measured $\Delta f$ shown in Fig.\,\fref{2}{2(a)} can be attributed to the occurrence of spin flop. Theoretically, the flopping behavior was a sudden action, which reflects a sharp jump in calculated results. Experimentally, the N\'eel vector deviated from the $c$ axis in CrPS$_{4}$ for $H\parallel c$\ucite{30} and the measured results show the $\Delta f$ gradually decreases below $H_{\rm sf}$. It should be noticed that the amplitude of the dip is dependent on the relative orientation of the crystal axes in CrPS$_{4}$ to the direction of cantilever $\hat{r}$. It becomes more pronounced when $a$ axis is mounted closer to $\hat{r}$ (as detailed in the Supporting Information). Nonetheless, the $H_{\rm sf}$ derived from the different sample is consistent. Above $H_{\rm sf}$, the calculated curve agrees well with the measurement results in S1 at 5\,K. The best fits give parameters of $M_0=1.65$\,kA/m, AV=0.0047 and $K=12.9$\,kJ/m$^3$. As shown in Fig.\,\fref{2}{2(b)}, the calculated $\Delta f$ rapidly increases with increasing $H$. A gradual decrease in its slope indicates that the canted AFM state undergoes a transition from initially metastable state to a steady state. Correspondingly, the spins rotate towards the external field, resulting in an increase in the total magnetic moment. $\Delta f$ shows a kink at $H=H_{\rm s}$, then eventually reaches saturation. For $H > H_{\rm s}$, the calculated $\Delta f$ shows a positive high-field asymptote behavior, where all the spins align in one direction as a field-polarized paramagnetic (FPP) type spin alignment and total magnetization satisfies $M_{\rm s}=2M_0$.

Figure~\fref{3}{3} summarized the $H$--$T$ diagram for $H\parallel c$ in CrPS$_{4}$ flake with different thickness. The diagram has been delineated into three regions. Below $H_{\rm sf}$, spins align as an A-AFM ground state. The decrease in $H_{\rm sf}$ was observed as the temperature increases, indicating the reduction in anisotropy. The region of $H_{\rm sf}<H<H_{\rm s}$, corresponding to the canted spin alignments, was denoted as canted AFM. For $H>H_{\rm s}$, the FPP region indicates a field-polarized paramagnetic-type spin alignments.

\vskip 4mm

\fl{4}\centerline{\includegraphics{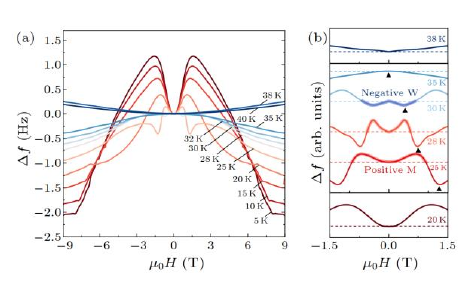}}

\vskip 2mm

\figcaption{7.5}{4}{Magnetic field dependence of $\Delta f$ in the case of $H \perp c$. (a) The magnetic field dependence of $\Delta f$ in S1 with $H \perp c$ from 5\,K to 40\,K. (b) The zoom-in $\Delta f$--$H$ curve at selected temperatures. The data has been normalized and shifted for clarify. The positive M shape and the negative W shape have been highlighted. The frequency shift at zero field is always zero. The dashed lines indicate the value $\Delta f=0$ at different temperatures. Black triangles mark the position of the valley.}

\medskip

To further investigate the temperature dependence of anisotropy in 2D CrPS$_{4}$, the DCM experiments have been conducted in S1 for $H\perp c$. Figure~\fref{4}{4(a)} shows field dependence of $\Delta f$ for $H\perp c$ in a temperature range from 5\,K to 40\,K. We firstly focus on the data at temperatures below 20\,K, where the $\Delta f$ at low field is positive. With further increasing $H$, the $\Delta f$ goes through a maximum, and reaches saturation with a change in slope for large $H$. The uniaxial model has been extended for the case of $H \perp c$ and $K>0$ (see the Supporting Information for  more details). The numerical results, indicating two macrospins gradually rotate to the external field with $\phi_{a_0}=\pi-\phi_{b_0}= \arcsin (\frac{M_0H}{K+2AM_0^2V})$ from the ac plane, agree well with our experiment data.

Now we turn to the evolution of metamagnetic phases above 20\,K. In Fig.\,\fref{4}{4(b)}, the curves have been normalized and shifted for clarity. As temperature increases, the shape of the $\Delta f$--$H$ curve gradually deviates from the M shape. Remarkably, a hump emerges right after the bump in the $\Delta f$ at around $H=1.3$\,T when $T$ reaches 25\,K. This change is induced by a relatively larger in-plane magnetization. In this case, the spins aligns close to $ab$ plane. The low-field $\Delta f$--$H$ curve becomes negative above 30\,K and maintains a W shape at low field until 35\,K. The negative signal indicates the dominance of in-plane anisotropy and the reorientation of the easy axis. The evolution of the curve shape shows that the spin reorientation is a continuous transition. Above 38\,K, the $\Delta f$ shows a quadratic field dependence. The upward parabola indicates that CrPS$_{4}$ enters a paramagnetic state where the in-plane shape anisotropy dominates.

\vskip 4mm

\fl{5}\centerline{\includegraphics{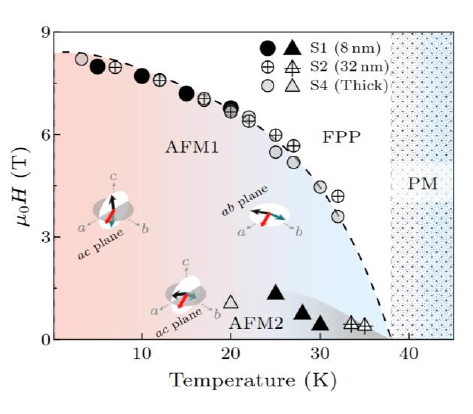}}

\vskip 2mm

\figcaption{7.5}{5}{Magnetic phase diagram of CrPS$_{4}$ with $H$ applied perpendicular to $c$ axis. The saturation fields (circles) and the reversal fields (triangles) are extracted from the $\Delta f$--$H$ curves in Figs.\,\fref{4}{4(a)} and \fref{4}{4(b)}, Figs.\,S2 and Fig.\,S4. The insets depict the corresponding spin alignments in different phases.}

\medskip
	
The $H$--$T$ phase diagram for $H\perp c$ has been summarized from different CrPS$_{4}$ samples. As shown in Fig.\,\fref{5}{5}, the phase diagram is roughly delineated as four regions. The FPP region represents the saturated magnetic phase with two macrospins aligning one direction. PM denots the paramagnetic phase of CrPS$_{4}$. The FPP-PM phase boundary is more distinct than the discussion before. AFM1 and AFM2 represent different cases for canted spin alignments, respectively. In AFM2, the spins align closer to the $a$ axis. With increasing magnetic field, the spins prefer to align closer to the $b$ axis (AFM1). The boundary between them is determined by the reversal field at the valley of $\Delta f$. Due to the different relative directions of crystal axes to the cantilever, the reversal fields vary for different samples. Nonetheless, the evolution of magnetic anisotropy induced by spin reorientation can be directly distinguished by performing DCM experiments.
	
The magnetic phase diagrams for $H\parallel c$ and $H\perp c$, constructed by DCM measurements in 2D CrPS$_{4}$, are basically consistent with that reported in bulk samples. Note that the alteration of spin orientation identified by commercial magnetometry is subtle and indirect. Specifically, the canted spin alignments and the evolution of spin reorientation in CrPS$_{4}$ can be only revealed by neutron scattering measurements, which is, however, unavailable for individual 2D particles. As discussed above, the effective magnetic spring constant $\Delta k$ is directly related to anisotropic magnetic susceptibility. The inherent dynamic property of DCM, on the other hand, makes it sensitive to quasi-static states. Moreover, the smallest detectable magnetotropic coefficient in our study can be equivalent to a magnetic moment $4.5\times 10^{6}\,\mu_{\rm B}$ at 1\,T (see Section S10 in the Supporting Information), which enables effective characterization of ultra weak magnetism even for single-layer samples. In this sense, DCM technique has the potential to investigate the exotic properties in 2D system, including skyrmions,\ucite{33,34,35} hidden orders,\ucite{36,37} and quantum critical phenomena.\ucite{38}
	
In summary, we have extended the high-precision DCM measurements to 2D antiferromagnets through a clean transfer technique. The effective magnetic spring constant $\Delta k$ has been measured in atomically thin CrPS$_{4}$ flake. Based on the field dependence of magnetic spring constant, the connection between the spin alignments and the $\Delta k$ of cantilever sensor is established to reveal the metamagnetic phase transitions for both $H\parallel c$ and $H\perp c$. The temperature evolution of the $\Delta f$--$H$ curve has been analyzed, from which we reveal the occurrence of spin reorientation. Our work pioneers magnetism measurement of a single atomically thin sample and opens a route to effectively explore novel phenomenon in 2D materials.
	
{\it Acknowledgements.} This work was supported by the National Key R\&D Program of China (Grant No.~2022YFA1602602), the National Natural Science Foundation of China (Grant Nos.~12122411 and 12474053), CAS Project for Young Scientists in Basic Research (Grant No.~YSBR-084), HFIPS Director's Fund (Grant Nos.~2023BR, YZJJ-GGZX-2022-03, and YZJJ202403-TS), HFIPS Director's Fud (Grant No.~BJPY2021B05), the Basic Research Program of the Chinese Academy of Sciences Based on Major Scientific Infrastructures (Grant No.~JZHKYPT-2021-08), and the High Magnetic Field Laboratory of Anhui Province (Grant No.~AHHM-FX-2020-02).

\end{multicols}
\end{document}